\documentclass[11pt]{article}
\usepackage{amsmath}
\usepackage{amssymb}
\usepackage{amsthm}
\usepackage{amsfonts}
\usepackage{comment}
\usepackage{color}
\usepackage{mathrsfs}
\usepackage{braket}
\usepackage{graphicx}
\usepackage[subrefformat=parens]{subcaption}
\usepackage{cite}
\usepackage{here}
\usepackage{tikz}
\usetikzlibrary{decorations.pathmorphing}
\usepackage{caption}
\usepackage{multirow}
\usepackage[labelfont=bf]{caption}
\usepackage{array} 
\usepackage{hyperref}
\usepackage{listings}
\usepackage{graphicx}

\usepackage[stable]{footmisc}
\makeatletter

\@addtoreset{equation}{section}
\makeatother

\begin{document}

%%%%%%%%%%%%%%%%%%%%%%%%%%
%\maketitle
\pagestyle{empty}

\title{\bf{Correspondence between quasinormal modes and grey-body factors in five-dimensional black holes}}

\date{}
\maketitle
\begin{center}
{\large 
Hyewon Han\footnote{dwhw101@dgu.ac.kr}, Bogeun Gwak\footnote{rasenis@dgu.ac.kr}
} \\
\vspace*{0.5cm}

{\it 
Department of Physics, Dongguk University, Seoul 04620, Republic of Korea
}

\end{center}

\vspace*{1.0cm}
\begin{abstract}

\end{abstract}
We investigate the correspondence between quasinormal modes and grey-body factors for the five-dimensional Schwarzschild--Tangherlini black hole. The quasinormal modes of gravitational perturbations are computed numerically using the continued fraction method. Particularly, we analyze the scalar, vector, and tensor types of perturbations, noting that the tensor type only exists in spacetimes with more than four dimensions. The grey-body factors are then calculated analytically via the correspondence, using data from the fundamental quasinormal mode and the first overtone. We then compute the grey-body factors independently using a numerical method and find that the results from the two methods coincide. This demonstrates that the correspondence holds with high accuracy for all three types of gravitational perturbations. Furthermore, our results extend the validity of the correspondence to the tensor type, which can only be tested in spacetimes with more than four dimensions.

\newpage
\baselineskip=18pt
\setcounter{page}{2}
\pagestyle{plain}
\baselineskip=18pt
\pagestyle{plain}
\setcounter{footnote}{0}

\section{Introduction}
A black hole, possessing an extremely strong gravitational field, is one of the most important and intriguing objects in the universe. The event horizon absorbs all matter and energy that crosses it, concealing its interior from any external observer. In general relativity, the Schwarzschild metric---which describes the simplest spherically symmetric black hole---is the first exact black hole solution discovered. The only conserved parameter is the mass of the black hole. Later, the charged Reissner--Nordström metric and the rotating Kerr metric were obtained. Various structures and physical properties, arising from black hole parameters and surrounding matter distributions, have been explored in theoretical and observational studies.

The interior of a black hole formed by gravitational collapse inevitably contains a singularity---a point where the curvature diverges. Numerous approaches have been proposed to address this high-curvature regime, among which higher-dimensional gravity has emerged as a promising avenue. In higher dimensions, black holes can possess novel horizon topologies \cite{Horowitz:1991cd} and rotational dynamics \cite{Myers:1986un}, leading to a wider variety of physical phenomena. Additionally, the gauge/gravity duality \cite{Maldacena:1997re,Gubser:1998bc,Witten:1998qj} linking $D$-dimensional quantum field theory to $(D+1)$-dimensional gravity theory motivated further study of black holes in higher dimensions. Particularly, five-dimensional spacetime---an extension with an extra spatial dimension---can represent other higher-dimensional spacetimes. Therefore, it has been regarded as a theoretical playground for exploring diverse gravitational phenomena.

The study of the interactions of a black hole with surrounding fields provides crucial insights into its strong gravitational regime. Two key quantities characterize the scattering and radiation processes in black hole spacetime: quasinormal modes and grey-body factors. A quasinormal mode is a damped oscillation with a characteristic complex frequency determined solely by the parameters of the black hole. This is defined by boundary conditions permitting purely ingoing waves at the event horizon and purely outgoing waves at spatial infinity. The ringdown phase of gravitational waves from merging binary black holes is well described by quasinormal modes, and can be used to estimate the parameters of the final remnant \cite{Dreyer:2003bv,Berti:2005ys,LIGOScientific:2020tif}. Additionally, quasinormal modes play a central role in numerous theoretical contexts, including superradiant instability \cite{Press:1972zz,Cardoso:2004nk,Cardoso:2004hs,Kodama:2009rq,Brito:2015oca} and strong cosmic censorship \cite{Cardoso:2017soq,Hod:2018lmi,Gwak:2018rba,Konoplya:2022kld,Singha:2022bvr,Davey:2024xvd}.

The grey-body factor quantifies the deviation of the radiation spectrum of a black hole from that of a perfect blackbody \cite{hawking1975particle,Hawking:1976de,Page:1976df}. This measures the fraction of radiation that escapes from the black hole into the asymptotic region by traversing a gravitational potential barrier. Owing to the symmetry of the scattering problem, we can consider waves incident from spatial infinity and compute the grey-body factor from their reflection and transmission amplitudes. Recent studies have shown that the grey-body factor can affect the ringdown spectral amplitude of gravitational waves \cite{Oshita:2022pkc,Oshita:2023cjz,Okabayashi:2024qbz}, highlighting its potential significance in gravitational wave detection and the test of gravity.

Quasinormal modes and grey-body factors are distinct spectral quantities obtained by solving field equations under different boundary conditions. Notably, a correspondence between them was recently established \cite{Konoplya:2024lir}. This correspondence employs the WKB approximation to relate quasinormal modes to grey-body factors. The WKB method matches asymptotic solutions of a wavelike equation with an effective potential to the Taylor expansion near the peak of the potential barrier. The method applies when the potential has a single peak with two turning points and asymptotically approaches constant values \cite{Schutz:1985km,Iyer:1986np,Konoplya:2003ii}. For black hole spacetimes with such potentials, the eikonal regime at large multipole numbers $l \gg 1$ is accurately described by the WKB method. Therefore, the correspondence derived from the WKB formula is exact in the eikonal limit. For small $l$, the correspondence is approximated with higher-order corrections and its accuracy requires verification. The correspondence was originally derived for spherically symmetric black holes and has been extended to rotating cases \cite{Konoplya:2024vuj}. More recently, it has been applied and tested in various four-dimensional models \cite{Skvortsova:2024msa,Dubinsky:2024vbn,Heidari:2024bbd,Malik:2024cgb,AraujoFilho:2025hkm,Konoplya:2025mvj,Lutfuoglu:2025hjy,Hamil:2025cms,Sajadi:2025kah,Tang:2025mkk,Pedrotti:2025idg,Lutfuoglu:2025ohb,Hamil:2025pte,Shi:2025gst,Bolokhov:2025lnt,Lutfuoglu:2025ldc,Ji:2025nlc}.

This study extends the analysis of the correspondence to higher dimensions, focusing on the gravitational perturbations of the five-dimensional Schwarzschild--Tangherlini black hole \cite{tangherlini1963schwarzschild}. In four dimensions, gravitational perturbations are classified into scalar and vector types, which share an identical quasinormal spectrum. In dimensions higher than four, a third type---the tensor mode---also appears. All three types possess distinct quasinormal spectra. For the five-dimensional black hole, we solved the equations for scalar-, vector-, and tensor-gravitational perturbations, and examined the correspondence between quasinormal modes and grey-body factors for each type. This is the first analysis of the correspondence for all three perturbation types in higher dimensions. The correspondence formula yields exact results in the eikonal limit $l \gg 1$; beyond this limit, it is approximate due to the asymptotic convergence of the WKB series. To test the approximate formula at low multipole numbers, we computed precise quasinormal modes and grey-body factors using numerical methods. The continued fraction method is one of the standard techniques for computing quasinormal modes of black holes \cite{Leaver:1985ax,Leaver:1990zz}. We obtained quasinormal frequencies for the fundamental mode and first overtone for each perturbation type, and determined the corresponding grey-body factors. By comparing these with numerically computed grey-body factors, we tested the validity of the correspondence for gravitational perturbations of the five-dimensional Schwarzschild--Tangherlini background. 
We used the signature $(-,+,+,+,+)$ and the units $c=G_5=1$ for the five-dimensional spacetime.

This paper is organized as follows. Section 2 presents the metric of the five-dimensional Schwarzschild--Tangherlini black hole. Section 3 solves the wavelike equations for the three perturbation types using the continued fraction method to obtain quasinormal frequencies. Section 4 computes grey-body factors from the correspondence and compares them with numerical results. Section 5 provides concluding remarks.

\section{Schwarzschild--Tangherlini black hole in five dimensions}
We aim to test the relationship between quasinormal modes and grey-body factors for scalar-, vector-, and tensor-gravitational perturbations in higher-dimensional black hole spacetimes. Higher-dimensional black holes arise as solutions to the Einstein field equations derived from the gravitational action
    \begin{align}
        I = \frac{1}{16\pi G_D} \int d^D x \sqrt{-g} \, \mathcal{R},
    \end{align}
where $D\ge4$ denotes the number of spacetime dimensions, $G_D$ is the $D$-dimensional Newton constant, $g$ is a determinant of the metric tensor, and $\mathcal{R}$ is the curvature scalar. Instead of considering arbitrary higher dimensions $D$, this study focuses specifically on the $D=5$ case. A spherically symmetric vacuum black hole is described by the Schwarzschild--Tangherlini solution \cite{tangherlini1963schwarzschild}, whose line element is given by
    \begin{align} \label{metric1}
        ds^2=-f(r) dt^2 +f^{-1}(r)dr^2+r^2 d \Omega_3^2 ,
    \end{align}
where $d\Omega_3^2=d\theta^2+\sin^2 \theta d\phi^2+\sin^2\theta \sin^2\phi d\chi^2$ denotes the line element of a unit three-sphere. The metric function $f(r)$ is given by
    \begin{align} 
        f(r)=1-\frac{M}{r^2},       
    \end{align}
and the mass parameter $M$ is related to the black hole mass $M_{BH} = \frac{3 \pi M}{8}$. From the metric \eqref{metric1}, the event horizon is located at $r_H$, such that $f(r_H)=0$, yielding
    \begin{align} 
        r_H= \sqrt{M} .     
    \end{align}
For notational simplicity, the metric function is rewritten as $f(r)=1-\left( \frac{r_H}{r}\right)^2$. The following sections consider gravitational perturbations in a fixed five-dimensional Schwarzschild--Tangherlini background and compute quasinormal modes and grey-body factors to examine their correspondence.

\section{Quasinormal modes of gravitational perturbations}
We first computed the quasinormal modes of gravitational perturbations. In typical cases with a small multipole number, the grey-body factor is determined primarily by the two dominant quasinormal modes: the fundamental mode ($n=0$) and the first overtone ($n=1$).

The master equation governing gravitational perturbations of spherically symmetric black holes in four and higher dimensions was derived by Kodama and Ishibashi \cite{Kodama:2003jz}. For $D\ge 5$, gravitational perturbations are classified into three types---scalar, vector, and tensor---based on their tensorial properties on the $(D-2)$-sphere. The scalar and vector types have four-dimensional counterparts, known as polar and axial modes, respectively, whereas tensor-gravitational perturbations occur only in higher dimensions. In four dimensions, scalar and vector perturbations are isospectral; that is, they share the same eigenvalue. In higher dimensions, no such isospectrality exists among the different perturbation types, so we computed the quasinormal frequencies and grey-body factors for all three.

\begin{figure}[h!] 
\noindent\begin{subfigure}[b]{0.33\textwidth}
    \centering
    \includegraphics[scale=0.55]{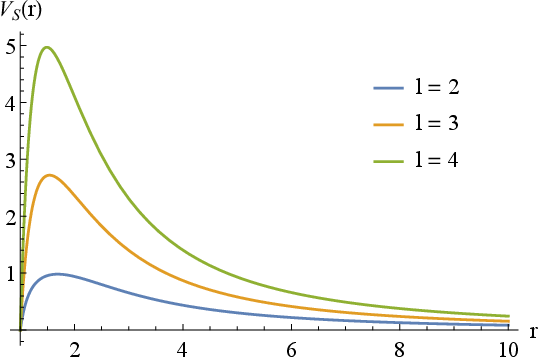}
\end{subfigure}%
\noindent\begin{subfigure}[b]{0.33\textwidth}
    \centering
    \includegraphics[scale=0.55]{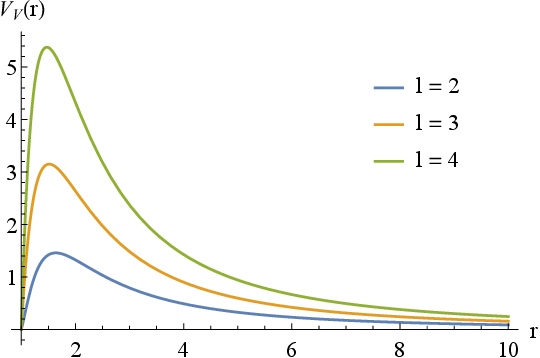}
\end{subfigure}%
\noindent\begin{subfigure}[b]{0.33\textwidth}
    \centering
    \includegraphics[scale=0.55]{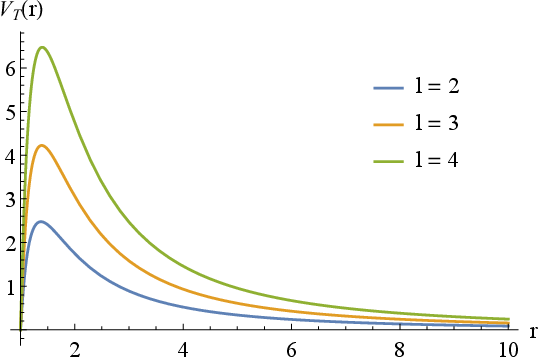}
\end{subfigure}%
\caption{The effective potentials for scalar (left), vector (center), and tensor (right) types of gravitational perturbations of the five-dimensional Schwarzschild--Tangherlini black hole ($r_H=1$) for $l=2$ (blue), $l=3$ (yellow), $l=4$ (green).} \label{fig1}
\end{figure}

In five dimensions, the Einstein field equations for gravitational perturbations of a Schwarzschild--Tangherlini black hole were reduced to a single second-order ordinary differential equation. For each tensorial type, this equation takes the form of a Schrödinger-like equation
    \begin{align} \label{waveeq}
        \frac{d^2 \Psi}{dr_*^2}+\left( \omega^2 - V(r) \right) \Psi=0,
    \end{align}
where $r_*(r)=\int f^{-1} dr$ is the tortoise coordinate, and the effective potentials are given by
    \begin{align}
        V_S(r)&= f(r) \left( \frac{\lambda^3}{r^2} +\frac{15 \lambda^2}{4r^2} +\frac{9 \lambda (\lambda-12) r_H^2}{4 r^4}+\frac{3(13\lambda-3) r_H^4}{r^6}+\frac{81 r_H^6}{r^8} \right) \left(\lambda+\frac{6 r_H^2}{r^2}\right)^{-2}, \label{Vs} \\
        V_V(r)&=f(r) \left( \frac{\lambda}{r^2}+\frac{15 f(r)}{4 r^2} - \frac{3r_H^2}{ r^4} \right), \label{VV} \\
        V_T(r)&=f(r) \left( \frac{\lambda}{r^2}+\frac{15 f(r)}{4 r^2} + \frac{6 r_H^2}{ r^4} \right), \label{VT}
    \end{align}
where $\lambda=l(l+2)-3$. The subscripts $S, V$, and $T$ refer to scalar, vector, and tensor perturbations, respectively, and $l \ge 2$ is the multipole number. The effective potentials for each type are shown in \autoref{fig1}. Each potential forms a positive barrier with a single peak, vanishing at the event horizon $r=r_H$ and spatial infinity $r=\infty$ corresponding to the boundaries. Consequently, the asymptotic solutions of the wave equation \eqref{waveeq} at these boundaries are easily obtained as
    \begin{align}
        \Psi = e^{\pm i \omega r_*}.
    \end{align}
Boundary conditions are imposed for quasinormal modes by requiring purely ingoing waves on the event horizon $r_*=-\infty$ and purely outgoing waves at spatial infinity $r_*=+\infty$, yielding asymptotic solutions
    \begin{align}
        \Psi &= e^{+ i \omega r_*}, \qquad r_* = + \infty, \label{bc1} \\
        \Psi &= e^{- i \omega r_*} ,\qquad r_* = - \infty, \label{bc2}
    \end{align}
where $\omega$ is quasinormal frequency with complex values, and the perturbations are assumed to have a time dependence $e^{-i\omega t}$.

To numerically determine $\omega$, we employed the continued fraction method, first introduced by Leaver \cite{Leaver:1985ax}, which remains one of the most accurate techniques for computing black hole quasinormal modes. For the five-dimensional Schwarzschild–Tangherlini black hole, we began by introducing the radial tortoise coordinate $r_*$
    \begin{align}
        r_*=\frac{1}{x}+\frac{1}{2x_H} \ln \left(\frac{x-x_H}{x_H} \right)-\frac{1}{2x_H} \ln \left(\frac{x+x_H}{x_H} \right),
    \end{align}
where we introduced $x=1/r$ and $x_H=1/r_H$. We considered the following ansatz \cite{Cardoso:2003vt,Cardoso:2003qd} for the wave function with boundary conditions \eqref{bc1}--\eqref{bc2}.
    \begin{align} \label{ansatz}
        \Psi=e^{i\omega/x} \left(\frac{x-x_H}{x_H} \right)^{-i\omega/(2x_H)} \left(\frac{x+x_H}{x_H} \right)^{-i\omega/(2x_H)} \sum^{\infty}_{k=0} a_k \left( \frac{x-x_H}{-x_H} \right)^k.
    \end{align}
Substituting this ansatz into the wave equation \eqref{waveeq} yields an $N$-term recurrence relation for the expansion coefficients $a_k$ as
    \begin{align} \label{recur}
        \sum^{\min(N-1,k+1)}_{j=0}c^{(N)}_{j,k}(\omega)  \, a_{k+1-j}=0, \qquad \mathrm{for} \ k=0,1,2,\cdots,
    \end{align}
where the coefficients $c^{(N)}_{j,k}(\omega)$ depend on the potential for each perturbation type. In the scalar-gravitational case, this relation involves eight terms, and explicit forms for $c^{(8)}_{j,k}$ are found as
    \begin{align} 
        c^{(8)}_{0,k}(\omega)=&-8 (1 + k) (6 + \lambda)^2 (1 + k - i r_H \omega), \nonumber\\
        c^{(8)}_{1,k}(\omega)=&4 (6 + \lambda) \left[12 + \lambda^2 + 5 k (6 + \lambda) + k^2 (78 + 5 \lambda) \right. \nonumber \\
        & \left. -i \left\{5(6+ \lambda)+2k(54+5\lambda) \right\}r_H \omega - 5 (6 + \lambda) r_H^2 \omega^2 \right], \nonumber \\
        c^{(8)}_{2,k}(\omega)=&-2952 - 2 \lambda (156 + \lambda) + 96 k (66 + 7 \lambda) - 16 k^2 ( 324 + 48\lambda + \lambda^2 ) \nonumber \\
        &- 32 i \left\{ 18(8 + \lambda)- k (270 + 45\lambda + \lambda^2) \right\} r_H \omega + 16 (6 + \lambda) (36 + \lambda) r_H^2 \omega^2, \nonumber \\
        c^{(8)}_{3,k}(\omega)=& 17604 + 1212 \lambda + \lambda^2 - 4 k (5580 + 384\lambda + \lambda^2) + 4 k^2 (1980 + 168\lambda + \lambda^2) \nonumber \\
        &+4 i \left\{ 5004 + 384\lambda + \lambda^2 - 2k (1836 + 168\lambda +\lambda^2) \right\} r_H \omega - 4 (1692 + 84 \lambda + \lambda^2) r_H^2 \omega^2, \nonumber \\
        c^{(8)}_{4,k}(\omega)=& 48 \left[ -762 - 25 \lambda + 22 k (30 + \lambda) - 6 k^2 (25 + \lambda)   \right. \nonumber \\
        & \left. -2i \left\{321+11\lambda-3k(49+2\lambda) \right\} r_H \omega + 6 (24 + \lambda) r_H^2 \omega^2 \right], \nonumber \\
        c^{(8)}_{5,k}(\omega)=&12 \left[2946 + 29 \lambda -4k(483+5\lambda) + 4 k^2 (81 + \lambda)  \right. \nonumber \\
        & \left. + 4 i \left\{483+5\lambda-2k(81+\lambda)\right\}r_H \omega - 4 (81 + \lambda) r_H^2 \omega^2 \right], \nonumber \\
        c^{(8)}_{6,k}(\omega)=&-72 \left[ 227 -120k+16k^2 +8i(15-4k) r_H \omega -16 r_H^2 \omega^2 \right], \nonumber \\
        c^{(8)}_{7,k}(\omega)=& 36 (9 - 2 k + 2 i r_H \omega)^2. \label{GE}
    \end{align} 
Using Gaussian elimination, the recurrence relation \eqref{recur} can be reduced to a three-term relation \cite{Leaver:1990zz, Zhidenko:2006rs,Konoplya:2011qq}. The number of terms decreases from $(p+1)$ to $p$ using the following relationship between the coefficients. 
    \begin{align} 
        c^{(p)}_{j,k}(\omega)&=c^{(p+1)}_{j,k}(\omega) \qquad \mathrm{for} \ j=0, \ \mathrm{or} \ (k+1) < p, \nonumber \\
        c^{(p)}_{j,k}(\omega)&=c^{(p+1)}_{j,k}(\omega) -\frac{c^{(p+1)}_{p,k}(\omega) \, c^{(p)}_{j-1,k-1}(\omega)}{c^{(p)}_{p-1,k-1}(\omega)}. \label{GaE}
    \end{align}
Applying this reduction successively from $p=7$ to $p=3$ yields the coefficients $c^{(3)}_{j,k}$ for the final three-term recurrence relation
    \begin{align} 
        \alpha_0 & a_1+\beta_0 a_0 =0,\\
        \alpha_k & a_{k+1}+\beta_k a_k + \gamma_k a_{k-1}=0, \qquad \mathrm{for} \ k>0,
    \end{align}
where $\alpha_k=c^{(3)}_{0, k}, \ \beta_k=c^{(3)}_{1, k}$, and $\gamma_k=c^{(3)}_{2, k}$. Here, the initial condition is fixed as $a_0=1$. The convergence condition for infinite series in the ansatz \eqref{ansatz} gives the equation for the quasinormal frequency.
    \begin{align} \label{n0}
        0=\beta_0 - \frac{\alpha_0 \gamma_1}{\beta_1-\frac{\alpha_1 \gamma_2}{\beta_2-\frac{\alpha_2 \gamma_3}{\beta_3- \cdots}}} \equiv \beta_0 - \frac{\alpha_0 \gamma_1}{\beta_1 -}\frac{\alpha_1 \gamma_2}{\beta_2 -}\frac{\alpha_2 \gamma_3}{\beta_3 -} \cdots.
    \end{align}
The frequency $\omega_0$ of the fundamental mode $n=0$ is obtained by solving it for specified values of $r_H$ and $l$. The frequency $\omega_1$ of the first overtone $n=1$ is found from the most stable root of the first inversion of the continued fraction, given by
    \begin{align} \label{n1}
        \beta_1 - \frac{\alpha_0 \gamma_1}{\beta_0} = \frac{\alpha_1 \gamma_2}{\beta_2 -}\frac{\alpha_2 \gamma_3}{\beta_3 -} \cdots.
    \end{align}
The resulting numerical values of $r_H \omega_n$ for $l=2,3,4$ are shown in \autoref{qnm1}. 

\begin{table}[h] 
    \centering  
    \begin{tabular} 
{ |>{\centering\arraybackslash}p{0.5cm}|
                  >{\centering\arraybackslash}p{3.5cm}|
                  >{\centering\arraybackslash}p{3.5cm}|
                  >{\centering\arraybackslash}p{3.5cm}|  }
 \hline
$n$ & $l=2$ & $l=3$ & $l=4$ \\
 \hline
 $0$   & $0.947739 - 0.256094 i$ & $1.605599-0.310960 i$ & $2.192404-0.329335 i$\\
 $1$   & $0.851234 - 0.821160 i$ & $1.510936-0.952776 i$ & $2.114851-0.999893 i$\\
 \hline
\end{tabular}
    \caption{The quasinormal frequencies $r_H \omega_n$ for scalar-gravitational perturbation of the five-dimensional Schwarzschild--Tangherlini black hole.}
\label{qnm1}
\end{table}

For vector-gravitational perturbations, the coefficients satisfy a four-term recurrence relation. These coefficients are explicit functions of the frequency and integer index $k$, given by
 \begin{align} 
        c^{(4)}_{0,k}(\omega)&-8 (1 + k) (1 + k - i r_H \omega), \nonumber\\
        c^{(4)}_{1,k}(\omega)&=4 \left[ -3 + 5 k^2 + 5 k (1 - 2 i r_H \omega) - 5 r_H \omega (i + r_H \omega) + \lambda\right], \nonumber \\
        c^{(4)}_{2,k}(\omega)&=70 - 16 (k - i r_H \omega)^2, \nonumber \\
        c^{(4)}_{3,k}(\omega)&= (-7 + 2 k - 2i r_H \omega) (5 + 2 k - 2 i r_H \omega).
    \end{align}
As with the scalar type, we obtained the coefficients of the three-term recurrence relation from Eq.~\eqref{GaE} for $p=3$ and numerically solved Eq.~\eqref{n0} and Eq.~\eqref{n1} to determine the quasinormal frequencies. 

\begin{table}[h] 
    \centering 
    \begin{tabular} 
{ |>{\centering\arraybackslash}p{0.5cm}|
                  >{\centering\arraybackslash}p{3.5cm}|
                  >{\centering\arraybackslash}p{3.5cm}|
                  >{\centering\arraybackslash}p{3.5cm}|  }
 \hline
$n$ & $l=2$ & $l=3$ & $l=4$ \\
 \hline
 $0$   & $1.134003 - 0.327523 i$ & $1.725363 - 0.333841 i$ & $2.280488 - 0.340017 i$\\
 $1$   & $0.947416 - 1.022040 i$ & $1.618094 - 1.019937 i$ & $2.200343 - 1.031684 i$\\
 \hline
\end{tabular}
    \caption{The quasinormal frequencies $r_H \omega_n$ for vector-gravitational perturbation of the five-dimensional Schwarzschild--Tangherlini black hole.} 
\label{qnm2}
\end{table} 

Finally, for tensor-gravitational perturbations, the coefficients of the four-term recurrence relation are obtained as
\begin{align} 
        c^{(4)}_{0,k}(\omega)&-8 (1 + k) (1 + k - i r_H \omega), \nonumber\\
        c^{(4)}_{1,k}(\omega)&=4 \left[6 + 5 k^2 + 5 k (1 - 2i r_H \omega) - 5 r_H \omega (i +r_H \omega) + \lambda\right], \nonumber \\
        c^{(4)}_{2,k}(\omega)&=-2 - 16 (k - i r_H \omega)^2, \nonumber \\
        c^{(4)}_{3,k}(\omega)&= (1 - 2 k + 2 i r_H \omega)^2.
    \end{align}
Using the same procedure, we computed the frequencies for the fundamental mode and first overtone. 

\begin{table}[h] 
    \centering 
    \begin{tabular} 
{ |>{\centering\arraybackslash}p{0.5cm}|
                  >{\centering\arraybackslash}p{3.5cm}|
                  >{\centering\arraybackslash}p{3.5cm}|
                  >{\centering\arraybackslash}p{3.5cm}|  }
 \hline
$n$ & $l=2$ & $l=3$ & $l=4$ \\
 \hline
 $0$   & $1.510567 - 0.357537 i$ & $2.007886 - 0.355802 i$ & $2.506291 - 0.354993 i
$\\
 $1$   & $1.392721 - 1.104557 i$ & $1.916954 - 1.085285 i$ & $2.432685 - 1.076372 i$\\
 \hline
\end{tabular}
    \caption{The quasinormal frequencies $r_H \omega_n$ for tensor-gravitational perturbation of the five-dimensional Schwarzschild--Tangherlini black hole.} 
\label{qnm3}
\end{table} 

We computed the quasinormal mode spectra for all three types of gravitational perturbations in the five-dimensional Schwarzschild--Tangherlini background. Each perturbation type exhibited a distinct spectrum. These results are used in the next section to examine the correspondence between quasinormal modes and grey-body factors.

\section{Correspondence between quasinormal modes and grey-body factors for  Schwarzschild--Tangherlini black hole in $D=5$}
A relation was established between quasinormal modes and grey-body factors, linking two spectral problems defined by different boundary conditions in black hole spacetimes. This relation was originally derived for spherically symmetric black holes in four dimensions \cite{Konoplya:2024lir}. In this study, we tested its validity by extending the analysis to a five-dimensional background. We calculated grey-body factors via correspondence, using quasinormal modes obtained with the continued fraction method, and compared them with direct numerical results.

The grey-body factor quantifies the transmission probability for radiation emitted from the black hole through its potential barrier. The symmetry allows us to assume an incident wave from spatial infinity and derive the same grey-body factor from the amplitude of the transmitted wave toward the event horizon. Therefore, we considered a wave equation
    \begin{align} \label{waveeq2}
        \frac{d^2 \Psi}{dr_*^2}+\left( \Omega^2 - V(r) \right) \Psi=0,
    \end{align}
with boundary conditions that allow purely ingoing waves at the horizon and both ingoing and outgoing waves at infinity. 
    \begin{align} \label{gbfbc}
        \Psi &= e^{- i \Omega r_*} + R  \, e^{+i \Omega r_*}, \qquad r_* = + \infty,  \\
        \Psi &= Te^{- i \Omega r_*} , \qquad \qquad\qquad r_* = - \infty,
    \end{align}
where $R$ and $T$ are called the reflection and transmission coefficients, respectively. The frequency $\Omega$ is real and continuous, unlike in complex and discrete quasinormal modes $\omega_n$. The grey-body factor is then determined by
    \begin{align}
        \Gamma(\Omega)=|T|^2 = 1- |R|^2,
    \end{align}
for a given value of $\Omega$.

The general WKB formula is written as an expansion around the eikonal limit, where the multipole number $l$ approaches infinity.
    \begin{align} \label{wkb}
        \Omega^2=& \,V_0 +A_2 (\mathcal{K}^2)+A_4 (\mathcal{K}^2)+A_6 (\mathcal{K}^2) + \cdots \nonumber \\
        &- i\mathcal{K} \sqrt{-2V_2} \left(1+ A_3 (\mathcal{K}^2)+A_5 (\mathcal{K}^2) + \cdots \, \right),
    \end{align}
where $V_0$ is the maximum value of the effective potential, $V_2$ is its second derivative at maximum, and $A_i$ is the higher-order correction terms for the eikonal formula \cite{Iyer:1986np,Konoplya:2003ii}. Then, the transmission and reflection coefficients can be expressed by
    \begin{align} \label{tc}
        |T|^2=\frac{1}{1+e^{2 \pi i \mathcal{K}}}=1- |R|^2,
    \end{align}
where $\mathcal{K}$ is a function of frequency $\Omega$ determined by Eq.~\eqref{wkb}. The WKB formula with higher-order corrections is also applied to the quasinormal frequency $\omega$, with the boundary conditions \eqref{bc1}--\eqref{bc2} yielding $\mathcal{K}=n+\frac{1}{2}$, where the overtone number is $n=0,1,2,\cdots$.

Using first-order WKB formulas in the eikonal approximation for quasinormal modes and grey-body factors, one can analytically obtain the exact transmission coefficient \eqref{tc} from the fundamental mode $\omega_0$. Beyond the eikonal limit, the sixth-order WKB method yields the relation incorporating correction terms with the fundamental mode $\omega_0$ and the first overtone $\omega_1$, improving accuracy.
    \begin{align}
         i\mathcal{K}=& \frac{\Omega^2 - \mathrm{Re}[\omega_0]^2}{4 \, \mathrm{Re}[\omega_0] \, \mathrm{Im}[\omega_0]} - \frac{\mathrm{Re}[\omega_0]-\mathrm{Re}[\omega_1]}{16 \, \mathrm{Im}[\omega_0]} \nonumber \\
        & +\frac{\Omega^2 - \mathrm{Re}[\omega_0]^2}{32 \, \mathrm{Re}[\omega_0] \, \mathrm{Im}[\omega_0]} \left(\frac{\left(\mathrm{Re}[\omega_0]-\mathrm{Re}[\omega_1]\right)^2}{4 \, \mathrm{Im}[\omega_0]^2} - \frac{3\mathrm{Im}[\omega_0]-\mathrm{Im}[\omega_1]}{3 \, \mathrm{Im}[\omega_0]} \right) \nonumber  \\
        &- \frac{\left(\Omega^2 - \mathrm{Re}[\omega_0]^2\right)^2}{16 \, \mathrm{Re}[\omega_0]^3 \, \mathrm{Im}[\omega_0]} \left(1+\frac{\mathrm{Re}[\omega_0]\left(\mathrm{Re}[\omega_0]-\mathrm{Re}[\omega_1]\right)}{4 \, \mathrm{Im}[\omega_0]^2} \right) \nonumber \\
        &+\frac{\left(\Omega^2 - \mathrm{Re}[\omega_0]^2\right)^3}{32 \, \mathrm{Re}[\omega_0]^5 \, \mathrm{Im}[\omega_0]} \left(1+\frac{\mathrm{Re}[\omega_0]\left(\mathrm{Re}[\omega_0]-\mathrm{Re}[\omega_1]\right)}{4 \, \mathrm{Im}[\omega_0]^2} \right. \nonumber \\
        & \left. \qquad \qquad \qquad + \mathrm{Re}[\omega_0]^2 \left( \frac{\left(\mathrm{Re}[\omega_0]-\mathrm{Re}[\omega_1]\right)^2}{16 \, \mathrm{Im}[\omega_0]^4} - \frac{3\mathrm{Im}[\omega_0]-\mathrm{Im}[\omega_1]}{12 \, \mathrm{Im}[\omega_0]} \right)\right) + \mathcal{O}(l^{-3}).\label{corr}
    \end{align}
When the WKB approach is valid, the first term on the right side yields accurate grey-body factors for large $l$. For smaller $l$, accuracy is improved by including the remaining terms of order $\mathcal{O}(l^{-1})$ and $\mathcal{O}(l^{-2})$. The correspondence \eqref{corr}, originally derived for a four-dimensional spherically symmetric black hole, can be applied to the higher-dimensional case without modification. We investigated whether the relationship between quasinormal modes and grey-body factors remains valid for the three types of gravitational perturbations in the five-dimensional Schwarzschild--Tangherlini spacetime.

\begin{figure}[h!] 
\noindent\begin{subfigure}[b]{0.5\textwidth}
    \centering
    \includegraphics[scale=0.8]{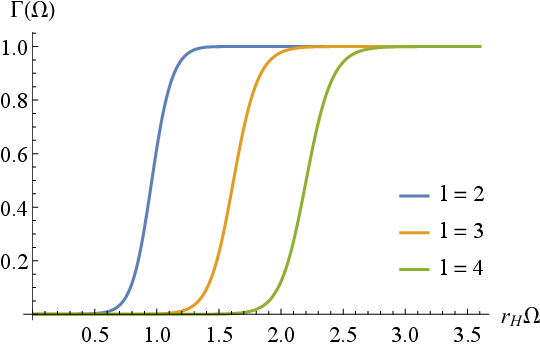}
\end{subfigure}%
\noindent\begin{subfigure}[b]{0.5\textwidth}
    \centering
    \includegraphics[scale=0.8]{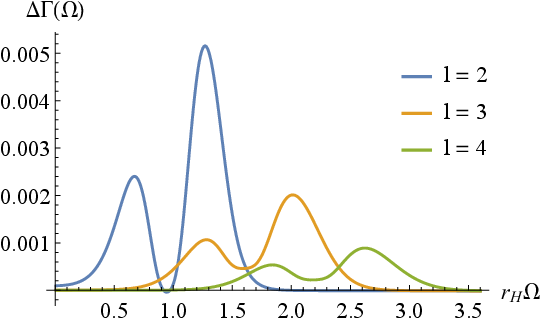}
\end{subfigure}
\caption{Left panel : the grey-body factors of scalar-gravitational perturbations obtained from the correspondence with quasinormal modes. Right panel: the differences between the results by correspondence and numerical method. The blue, yellow, and green lines are for $l=2,3,4$ respectively.} \label{fig2}
\end{figure}

\begin{figure}[h!] 
\noindent\begin{subfigure}[b]{0.5\textwidth}
    \centering
    \includegraphics[scale=0.8]{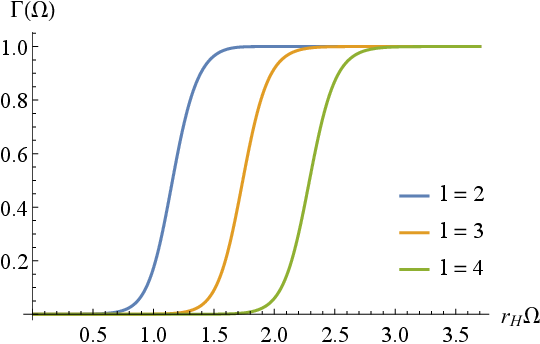}
\end{subfigure}%
\noindent\begin{subfigure}[b]{0.5\textwidth}
    \centering
    \includegraphics[scale=0.8]{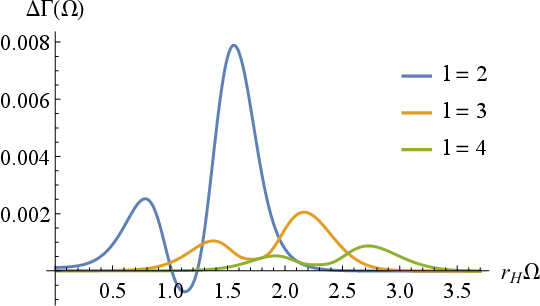}
\end{subfigure}
\caption{Left panel : the grey-body factors of vector-gravitational perturbations obtained from the correspondence with quasinormal modes. Right panel: the differences between the results by correspondence and numerical method. The blue, yellow, and green lines are for $l=2,3,4$ respectively.} \label{fig3}
\end{figure}

\begin{figure}[h!] 
\noindent\begin{subfigure}[b]{0.5\textwidth}
    \centering
    \includegraphics[scale=0.8]{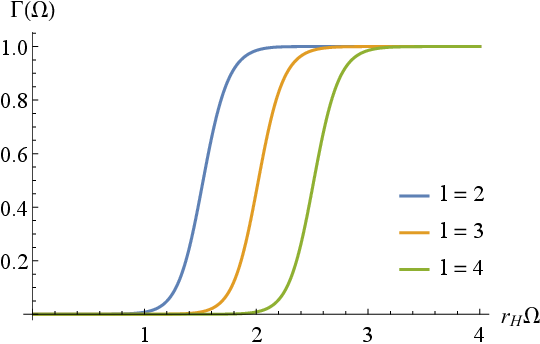}
\end{subfigure}%
\noindent\begin{subfigure}[b]{0.5\textwidth}
    \centering
    \includegraphics[scale=0.8]{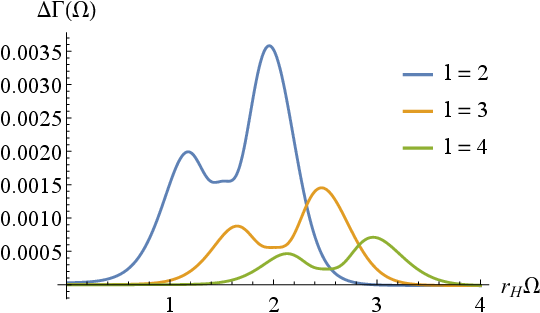}
\end{subfigure}
\caption{Left panel : the grey-body factors of tensor-gravitational perturbations obtained from the correspondence with quasinormal modes. Right panel: the differences between the results by correspondence and numerical method. The blue, yellow, and green lines are for $l=2,3,4$ respectively.} \label{fig4}
\end{figure}

We first computed the grey-body factors by substituting the quasinormal frequencies $\omega_0$ and $\omega_1$ from \autoref{qnm1}--\ref{qnm3} into Eq.~\eqref{corr}. The grey-body factors for scalar, vector, and tensor perturbations are plotted for a real frequency $\Omega$ in the left panels of \autoref{fig2}--\ref{fig4}, respectively. They all show gradual increases from zero to one as the wave energy rises.

To obtain accurate reference values for the grey-body factors, we numerically solved Eq.~\eqref{waveeq2} using \texttt{GrayHawk}, a recently published code for computing grey-body factors of spherically symmetric black holes \cite{Calza:2025whq}, modified for the five-dimensional Schwarzschild--Tangherlini background. For the four-dimensional Schwarzschild black hole, results from this tool align with the correspondence values reported in \cite{Konoplya:2024lir}. Motivated by this agreement, we extended the analysis to five dimensions to test the correspondence. In our modified version, the sampling interval was refined to improve resolution, addressing the issue of insufficient sampling density, particularly in the high-frequency regime.

The differences between the numerical results and the values predicted by correspondence \eqref{corr} are shown in the right panels of \autoref{fig2}--\ref{fig4}. For all three types of gravitational perturbations, these differences are small, indicating that the correspondence remains valid in five dimensions. As the multipole number $l$ increases, the differences diminish, remaining below $0.001$ for $l=4$. Therefore, the correspondence becomes more accurate for large $l$, and is expected to be exact in the eikonal limit.

\begin{figure}[h!] 
\noindent\begin{subfigure}[b]{0.33\textwidth}
    \centering
    \includegraphics[scale=0.55]{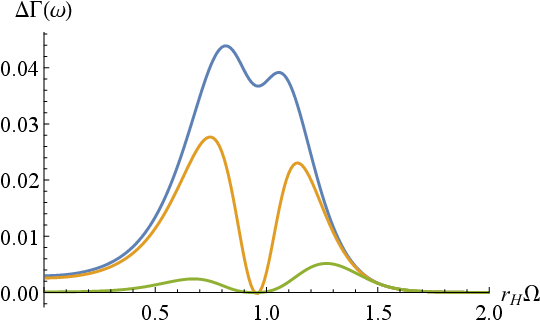}
\end{subfigure}%
\noindent\begin{subfigure}[b]{0.33\textwidth}
    \centering
    \includegraphics[scale=0.55]{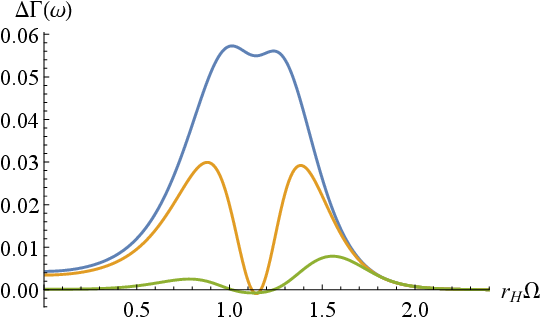}
\end{subfigure}%
\noindent\begin{subfigure}[b]{0.33\textwidth}
    \centering
    \includegraphics[scale=0.55]{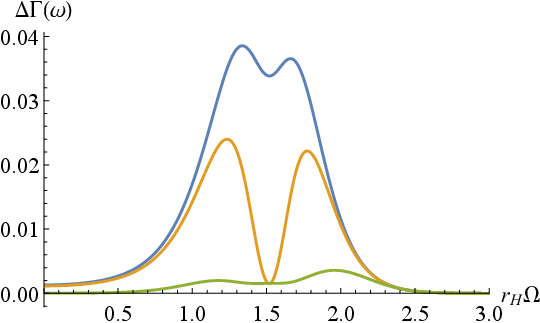}
\end{subfigure}%
\caption{Differences between the grey-body factors obtained using correspondence and numerical methods for scalar (left), vector (center), and tensor (right) types of perturbations for $l=2$. The blue line is for the eikonal formula, the yellow line for the first-order beyond eikonal formula, and the green line for the complete formula containing the corrections of the order $\mathcal{O}(l^{-2})$.} \label{fig5}
\end{figure}

Furthermore, we examined the effect of corrections to the eikonal formula on the relationship between quasinormal modes and grey-body factors, focusing on the lowest multipole number $l=2$. In Eq.~\eqref{corr}, the first term corresponds to the eikonal formula, the second term represents the first-order correction derived from the second-order WKB approximation, and the full sixth-order WKB expression includes corrections of the order $\mathcal{O}(l^{-2})$. The differences between the grey-body factors from each formula and the accurate values are plotted in \autoref{fig5}. The results demonstrate that the complete \eqref{corr}, incorporating beyond-eikonal corrections, provides markedly higher accuracy, as indicated by the green curves. 

We confirmed that the grey-body factors for all types of gravitational perturbations in the five-dimensional Schwarzschild--Tangherlini black hole are accurately obtained by the analytical expression \eqref{corr} derived from the quasinormal modes. This accuracy arises from the form of the effective potential in the Schwarzschild--Tangherlini spacetime, for which the WKB approximation underlying the correspondence performs well \cite{Konoplya:2003ii}. We therefore expect the correspondence to remain valid for spherically symmetric black holes in dimensions higher than five.

\section{Conclusions}

In this work, the correspondence between quasinormal modes and grey-body factors was analyzed in five dimensions. Our analysis focused on gravitational perturbations of the Schwarzschild--Tangherlini spacetime. The correspondence, as derived via the sixth-order WKB method, is exact in the eikonal limit $l \gg 1$. To examine its validity for small multipole numbers, we computed accurate values of quasinormal modes and grey-body factors using numerical methods and then compared these with the results predicted by the correspondence. 

Since distinct types of gravitational perturbations are not isospectral in higher dimensions, we considered scalar, vector, and tensor perturbations separately. For each type, we solved the wave equation using the continued fraction method and obtained the two dominant quasinormal modes, $\omega_0$ and $\omega_1$. These modes were then used to determine the grey-body factors via an analytical formula. Additionally, we solved the equation numerically to obtain grey-body factors and compared them with those obtained by the correspondence. For all perturbation types, the results show high accuracy of relationship between quasinormal modes and grey-body factors for the lowest multipole numbers $l=2,3,4$. This alignment is attributed to the favorable performance of the WKB method for spherically symmetric black holes in five dimensions. Therefore, we expect the correspondence to remain valid for such spacetimes in dimensions higher than five.

\vspace{10pt} 

\noindent{\bf Acknowledgments}

\noindent This research was supported by Basic Science Research Program through the National Research Foundation of Korea (NRF) funded by the Ministry of Education (NRF-2022R1I1A2063176) and the Dongguk University Research Fund of 2025.\\

\bibliographystyle{bibstyle}
\bibliography{ref}

\providecommand{\href}[2]{#2}\begingroup\raggedright\begin{thebibliography}{10}

\bibitem{Horowitz:1991cd}
G.~T. Horowitz and A.~Strominger, \emph{{Black strings and P-branes}}, \href{https://doi.org/10.1016/0550-3213(91)90440-9}{\emph{Nucl. Phys. B} {\bfseries 360} (1991) 197}.

\bibitem{Myers:1986un}
R.~C. Myers and M.~J. Perry, \emph{{Black Holes in Higher Dimensional Space-Times}}, \href{https://doi.org/10.1016/0003-4916(86)90186-7}{\emph{Annals Phys.} {\bfseries 172} (1986) 304}.

\bibitem{Maldacena:1997re}
J.~M. Maldacena, \emph{{The Large $N$ limit of superconformal field theories and supergravity}}, \href{https://doi.org/10.4310/ATMP.1998.v2.n2.a1}{\emph{Adv. Theor. Math. Phys.} {\bfseries 2} (1998) 231} [\href{https://arxiv.org/abs/hep-th/9711200}{{\ttfamily hep-th/9711200}}].

\bibitem{Gubser:1998bc}
S.~S. Gubser, I.~R. Klebanov and A.~M. Polyakov, \emph{{Gauge theory correlators from noncritical string theory}}, \href{https://doi.org/10.1016/S0370-2693(98)00377-3}{\emph{Phys. Lett. B} {\bfseries 428} (1998) 105} [\href{https://arxiv.org/abs/hep-th/9802109}{{\ttfamily hep-th/9802109}}].

\bibitem{Witten:1998qj}
E.~Witten, \emph{{Anti de Sitter space and holography}}, \href{https://doi.org/10.4310/ATMP.1998.v2.n2.a2}{\emph{Adv. Theor. Math. Phys.} {\bfseries 2} (1998) 253} [\href{https://arxiv.org/abs/hep-th/9802150}{{\ttfamily hep-th/9802150}}].

\bibitem{Dreyer:2003bv}
O.~Dreyer, B.~J. Kelly, B.~Krishnan, L.~S. Finn, D.~Garrison and R.~Lopez-Aleman, \emph{{Black hole spectroscopy: Testing general relativity through gravitational wave observations}}, \href{https://doi.org/10.1088/0264-9381/21/4/003}{\emph{Class. Quant. Grav.} {\bfseries 21} (2004) 787} [\href{https://arxiv.org/abs/gr-qc/0309007}{{\ttfamily gr-qc/0309007}}].

\bibitem{Berti:2005ys}
E.~Berti, V.~Cardoso and C.~M. Will, \emph{{On gravitational-wave spectroscopy of massive black holes with the space interferometer LISA}}, \href{https://doi.org/10.1103/PhysRevD.73.064030}{\emph{Phys. Rev. D} {\bfseries 73} (2006) 064030} [\href{https://arxiv.org/abs/gr-qc/0512160}{{\ttfamily gr-qc/0512160}}].

\bibitem{LIGOScientific:2020tif}
{\scshape LIGO Scientific, Virgo} collaboration, \emph{{Tests of general relativity with binary black holes from the second LIGO-Virgo gravitational-wave transient catalog}}, \href{https://doi.org/10.1103/PhysRevD.103.122002}{\emph{Phys. Rev. D} {\bfseries 103} (2021) 122002} [\href{https://arxiv.org/abs/2010.14529}{{\ttfamily 2010.14529}}].

\bibitem{Press:1972zz}
W.~H. Press and S.~A. Teukolsky, \emph{{Floating Orbits, Superradiant Scattering and the Black-hole Bomb}}, \href{https://doi.org/10.1038/238211a0}{\emph{Nature} {\bfseries 238} (1972) 211}.

\bibitem{Cardoso:2004nk}
V.~Cardoso, O.~J.~C. Dias, J.~P.~S. Lemos and S.~Yoshida, \emph{{The Black hole bomb and superradiant instabilities}}, \href{https://doi.org/10.1103/PhysRevD.70.049903}{\emph{Phys. Rev. D} {\bfseries 70} (2004) 044039} [\href{https://arxiv.org/abs/hep-th/0404096}{{\ttfamily hep-th/0404096}}].

\bibitem{Cardoso:2004hs}
V.~Cardoso and O.~J.~C. Dias, \emph{{Small Kerr-anti-de Sitter black holes are unstable}}, \href{https://doi.org/10.1103/PhysRevD.70.084011}{\emph{Phys. Rev. D} {\bfseries 70} (2004) 084011} [\href{https://arxiv.org/abs/hep-th/0405006}{{\ttfamily hep-th/0405006}}].

\bibitem{Kodama:2009rq}
H.~Kodama, R.~A. Konoplya and A.~Zhidenko, \emph{{Gravitational instability of simply rotating AdS black holes in higher dimensions}}, \href{https://doi.org/10.1103/PhysRevD.79.044003}{\emph{Phys. Rev. D} {\bfseries 79} (2009) 044003} [\href{https://arxiv.org/abs/0812.0445}{{\ttfamily 0812.0445}}].

\bibitem{Brito:2015oca}
R.~Brito, V.~Cardoso and P.~Pani, \emph{{Superradiance}: {New Frontiers in Black Hole Physics}}, \href{https://doi.org/10.1007/978-3-319-19000-6}{\emph{Lect. Notes Phys.} {\bfseries 906} (2015) pp.1} [\href{https://arxiv.org/abs/1501.06570}{{\ttfamily 1501.06570}}].

\bibitem{Cardoso:2017soq}
V.~Cardoso, J.~L. Costa, K.~Destounis, P.~Hintz and A.~Jansen, \emph{{Quasinormal modes and Strong Cosmic Censorship}}, \href{https://doi.org/10.1103/PhysRevLett.120.031103}{\emph{Phys. Rev. Lett.} {\bfseries 120} (2018) 031103} [\href{https://arxiv.org/abs/1711.10502}{{\ttfamily 1711.10502}}].

\bibitem{Hod:2018lmi}
S.~Hod, \emph{{Quasinormal modes and strong cosmic censorship in near-extremal Kerr\textendash{}Newman\textendash{}de Sitter black-hole spacetimes}}, \href{https://doi.org/10.1016/j.physletb.2018.03.020}{\emph{Phys. Lett. B} {\bfseries 780} (2018) 221} [\href{https://arxiv.org/abs/1803.05443}{{\ttfamily 1803.05443}}].

\bibitem{Gwak:2018rba}
B.~Gwak, \emph{{Strong Cosmic Censorship under Quasinormal Modes of Non-Minimally Coupled Massive Scalar Field}}, \href{https://doi.org/10.1140/epjc/s10052-019-7283-5}{\emph{Eur. Phys. J. C} {\bfseries 79} (2019) 767} [\href{https://arxiv.org/abs/1812.04923}{{\ttfamily 1812.04923}}].

\bibitem{Konoplya:2022kld}
R.~A. Konoplya and A.~Zhidenko, \emph{{How general is the strong cosmic censorship bound for quasinormal modes?}}, \href{https://doi.org/10.1088/1475-7516/2022/11/028}{\emph{JCAP} {\bfseries 11} (2022) 028} [\href{https://arxiv.org/abs/2210.04314}{{\ttfamily 2210.04314}}].

\bibitem{Singha:2022bvr}
C.~Singha, S.~Chakraborty and N.~Dadhich, \emph{{Strong cosmic censorship conjecture for a charged BTZ black hole}}, \href{https://doi.org/10.1007/JHEP06(2022)028}{\emph{JHEP} {\bfseries 06} (2022) 028} [\href{https://arxiv.org/abs/2203.07708}{{\ttfamily 2203.07708}}].

\bibitem{Davey:2024xvd}
A.~Davey, O.~J.~C. Dias and D.~S. Gil, \emph{{Strong Cosmic Censorship in Kerr-Newman-de Sitter}}, \href{https://doi.org/10.1007/JHEP07(2024)113}{\emph{JHEP} {\bfseries 07} (2024) 113} [\href{https://arxiv.org/abs/2404.03724}{{\ttfamily 2404.03724}}].

\bibitem{hawking1975particle}
S.~W. Hawking, \emph{{Particle Creation by Black Holes}}, \href{https://doi.org/10.1007/BF02345020}{\emph{Commun. Math. Phys.} {\bfseries 43} (1975) 199}.

\bibitem{Hawking:1976de}
S.~W. Hawking, \emph{{Black Holes and Thermodynamics}}, \href{https://doi.org/10.1103/PhysRevD.13.191}{\emph{Phys. Rev. D} {\bfseries 13} (1976) 191}.

\bibitem{Page:1976df}
D.~N. Page, \emph{{Particle Emission Rates from a Black Hole: Massless Particles from an Uncharged, Nonrotating Hole}}, \href{https://doi.org/10.1103/PhysRevD.13.198}{\emph{Phys. Rev. D} {\bfseries 13} (1976) 198}.

\bibitem{Oshita:2022pkc}
N.~Oshita, \emph{{Thermal ringdown of a Kerr black hole: overtone excitation, Fermi-Dirac statistics and greybody factor}}, \href{https://doi.org/10.1088/1475-7516/2023/04/013}{\emph{JCAP} {\bfseries 04} (2023) 013} [\href{https://arxiv.org/abs/2208.02923}{{\ttfamily 2208.02923}}].

\bibitem{Oshita:2023cjz}
N.~Oshita, \emph{{Greybody factors imprinted on black hole ringdowns: An alternative to superposed quasinormal modes}}, \href{https://doi.org/10.1103/PhysRevD.109.104028}{\emph{Phys. Rev. D} {\bfseries 109} (2024) 104028} [\href{https://arxiv.org/abs/2309.05725}{{\ttfamily 2309.05725}}].

\bibitem{Okabayashi:2024qbz}
K.~Okabayashi and N.~Oshita, \emph{{Greybody factors imprinted on black hole ringdowns. II. Merging binary black holes}}, \href{https://doi.org/10.1103/PhysRevD.110.064086}{\emph{Phys. Rev. D} {\bfseries 110} (2024) 064086} [\href{https://arxiv.org/abs/2403.17487}{{\ttfamily 2403.17487}}].

\bibitem{Konoplya:2024lir}
R.~A. Konoplya and A.~Zhidenko, \emph{{Correspondence between grey-body factors and quasinormal modes}}, \href{https://doi.org/10.1088/1475-7516/2024/09/068}{\emph{JCAP} {\bfseries 09} (2024) 068} [\href{https://arxiv.org/abs/2406.11694}{{\ttfamily 2406.11694}}].

\bibitem{Schutz:1985km}
B.~F. Schutz and C.~M. Will, \emph{{BLACK HOLE NORMAL MODES: A SEMIANALYTIC APPROACH}}, \href{https://doi.org/10.1086/184453}{\emph{Astrophys. J. Lett.} {\bfseries 291} (1985) L33}.

\bibitem{Iyer:1986np}
S.~Iyer and C.~M. Will, \emph{{Black Hole Normal Modes: A {WKB} Approach. 1. Foundations and Application of a Higher Order {WKB} Analysis of Potential Barrier Scattering}}, \href{https://doi.org/10.1103/PhysRevD.35.3621}{\emph{Phys. Rev. D} {\bfseries 35} (1987) 3621}.

\bibitem{Konoplya:2003ii}
R.~A. Konoplya, \emph{{Quasinormal behavior of the d-dimensional Schwarzschild black hole and higher order WKB approach}}, \href{https://doi.org/10.1103/PhysRevD.68.024018}{\emph{Phys. Rev. D} {\bfseries 68} (2003) 024018} [\href{https://arxiv.org/abs/gr-qc/0303052}{{\ttfamily gr-qc/0303052}}].

\bibitem{Konoplya:2024vuj}
R.~A. Konoplya and A.~Zhidenko, \emph{{Correspondence between grey-body factors and quasinormal frequencies for rotating black holes}}, \href{https://doi.org/10.1016/j.physletb.2025.139288}{\emph{Phys. Lett. B} {\bfseries 861} (2025) 139288} [\href{https://arxiv.org/abs/2408.11162}{{\ttfamily 2408.11162}}].

\bibitem{Skvortsova:2024msa}
M.~Skvortsova, \emph{{Quantum corrected black holes: testing the correspondence between grey-body factors and quasinormal modes}},  \href{https://arxiv.org/abs/2411.06007}{{\ttfamily 2411.06007}}.

\bibitem{Dubinsky:2024vbn}
A.~Dubinsky, \emph{{Grey-body factors for gravitational and electromagnetic perturbations around Gibbons-Maeda-Garfinkle-Horovits-Strominger black holes}},  \href{https://arxiv.org/abs/2412.00625}{{\ttfamily 2412.00625}}.

\bibitem{Heidari:2024bbd}
N.~Heidari, A.~A. Ara{\'u}jo~Filho, V.~Vertogradov and A.~{\"O}vg{\"u}n, \emph{{Black hole with a de Sitter core: classical and quantum features}},  \href{https://arxiv.org/abs/2412.05072}{{\ttfamily 2412.05072}}.

\bibitem{Malik:2024cgb}
Z.~Malik, \emph{{Correspondence between quasinormal modes and grey-body factors for massive fields in Schwarzschild-de~Sitter spacetime}}, \href{https://doi.org/10.1088/1475-7516/2025/04/042}{\emph{JCAP} {\bfseries 04} (2025) 042} [\href{https://arxiv.org/abs/2412.19443}{{\ttfamily 2412.19443}}].

\bibitem{AraujoFilho:2025hkm}
A.~A. Ara{\'u}jo~Filho, \emph{{How does non-metricity affect particle creation and evaporation in bumblebee gravity?}}, \href{https://doi.org/10.1088/1475-7516/2025/06/026}{\emph{JCAP} {\bfseries 06} (2025) 026} [\href{https://arxiv.org/abs/2501.00927}{{\ttfamily 2501.00927}}].

\bibitem{Konoplya:2025mvj}
R.~A. Konoplya, A.~Khrabustovskyi, J.~K{\v{r}}{\'\i}{\v{z}} and A.~Zhidenko, \emph{{Quasinormal ringing and shadows of black holes and wormholes in dark matter-inspired Weyl gravity}}, \href{https://doi.org/10.1088/1475-7516/2025/04/062}{\emph{JCAP} {\bfseries 04} (2025) 062} [\href{https://arxiv.org/abs/2501.16134}{{\ttfamily 2501.16134}}].

\bibitem{Lutfuoglu:2025hjy}
B.~C. L{\"u}tf{\"u}o{\u{g}}lu, \emph{{Long-lived quasinormal modes and gray-body factors of black holes and wormholes in dark matter inspired Weyl gravity}}, \href{https://doi.org/10.1140/epjc/s10052-025-14210-0}{\emph{Eur. Phys. J. C} {\bfseries 85} (2025) 486} [\href{https://arxiv.org/abs/2503.16087}{{\ttfamily 2503.16087}}].

\bibitem{Hamil:2025cms}
B.~Hamil and B.~C. L{\"u}tf{\"u}o{\u{g}}lu, \emph{{Nonlinear Magnetically Charged Black Holes with Phantom Global Monopoles: Thermodynamics, Geodesics, Quasinormal Modes, and Grey-Body Factors}},  \href{https://arxiv.org/abs/2503.17474}{{\ttfamily 2503.17474}}.

\bibitem{Sajadi:2025kah}
S.~N. Sajadi, S.~Ponglertsakul and D.~J. Gogoi, \emph{{Physical Properties of Black Hole Solution in Einstein-Bel-Robinson Gravity}},  \href{https://arxiv.org/abs/2503.18289}{{\ttfamily 2503.18289}}.

\bibitem{Tang:2025mkk}
C.~Tang, Y.~Ling and Q.-Q. Jiang, \emph{{Correspondence between grey-body factors and quasinormal modes for regular black holes with sub-Planckian curvature}},  \href{https://arxiv.org/abs/2503.21597}{{\ttfamily 2503.21597}}.

\bibitem{Pedrotti:2025idg}
D.~Pedrotti and M.~Calz{\`a}, \emph{{Trinity of black hole correspondences: Shadows, quasinormal modes, graybody factors, and cautionary remarks}}, \href{https://doi.org/10.1103/1q35-mjjz}{\emph{Phys. Rev. D} {\bfseries 111} (2025) 124056} [\href{https://arxiv.org/abs/2504.01909}{{\ttfamily 2504.01909}}].

\bibitem{Lutfuoglu:2025ohb}
B.~C. L{\"u}tf{\"u}o{\u{g}}lu, \emph{{Quasinormal Modes and Gray-Body Factors for Gravitational Perturbations in Asymptotically Safe Gravity}},  \href{https://arxiv.org/abs/2505.06966}{{\ttfamily 2505.06966}}.

\bibitem{Hamil:2025pte}
B.~Hamil, A.~Al-Badawi and B.~C. L{\"u}tf{\"u}o{\u{g}}lu, \emph{{Geodesics and scalar perturbations of Schwarzschild black holes embedded in a Dehnen-type dark matter halo with quintessence}},  \href{https://arxiv.org/abs/2505.18611}{{\ttfamily 2505.18611}}.

\bibitem{Shi:2025gst}
Q.-L. Shi, R.~Wang, W.~Xiong and P.-C. Li, \emph{{Quasinormal modes and grey-body factors of axial gravitational perturbations of regular black holes in asymptotically safe gravity}},  \href{https://arxiv.org/abs/2506.16217}{{\ttfamily 2506.16217}}.

\bibitem{Bolokhov:2025lnt}
S.~V. Bolokhov and M.~Skvortsova, \emph{{Gravitational Quasinormal Modes and Grey-Body Factors of Bonanno-Reuter Regular Black Holes}}, {\emph{Theor. Phys.} {\bfseries 1} (2025) 3} [\href{https://arxiv.org/abs/2507.07196}{{\ttfamily 2507.07196}}].

\bibitem{Lutfuoglu:2025ldc}
B.~C. L{\"u}tf{\"u}o{\u{g}}lu, \emph{{Black Holes in Proca-Gauss-Bonnet Gravity with Primary Hair: Particle Motion, Shadows, and Grey-Body Factors}}, {\emph{Theor. Phys.} {\bfseries 1} (2025) 4} [\href{https://arxiv.org/abs/2507.09246}{{\ttfamily 2507.09246}}].

\bibitem{Ji:2025nlc}
Y.~Ji, \emph{{Quasinormal modes and Hawking radiation of a non-minimal Einstein-Yang-Mills regular black hole}}, \href{https://doi.org/10.1088/1402-4896/adeede}{\emph{Phys. Scripta} {\bfseries 100} (2025) 085001}.

\bibitem{tangherlini1963schwarzschild}
F.~R. Tangherlini, \emph{Schwarzschild field in n dimensions and the dimensionality of space problem}, {\emph{Il Nuovo Cimento (1955-1965)} {\bfseries 27} (1963) 636}.

\bibitem{Leaver:1985ax}
E.~W. Leaver, \emph{{An Analytic representation for the quasi normal modes of Kerr black holes}}, \href{https://doi.org/10.1098/rspa.1985.0119}{\emph{Proc. Roy. Soc. Lond. A} {\bfseries 402} (1985) 285}.

\bibitem{Leaver:1990zz}
E.~W. Leaver, \emph{{Quasinormal modes of Reissner-Nordstrom black holes}}, \href{https://doi.org/10.1103/PhysRevD.41.2986}{\emph{Phys. Rev. D} {\bfseries 41} (1990) 2986}.

\bibitem{Kodama:2003jz}
H.~Kodama and A.~Ishibashi, \emph{{A Master equation for gravitational perturbations of maximally symmetric black holes in higher dimensions}}, \href{https://doi.org/10.1143/PTP.110.701}{\emph{Prog. Theor. Phys.} {\bfseries 110} (2003) 701} [\href{https://arxiv.org/abs/hep-th/0305147}{{\ttfamily hep-th/0305147}}].

\bibitem{Cardoso:2003vt}
V.~Cardoso, J.~P.~S. Lemos and S.~Yoshida, \emph{{Quasinormal modes of Schwarzschild black holes in four-dimensions and higher dimensions}}, \href{https://doi.org/10.1103/PhysRevD.69.044004}{\emph{Phys. Rev. D} {\bfseries 69} (2004) 044004} [\href{https://arxiv.org/abs/gr-qc/0309112}{{\ttfamily gr-qc/0309112}}].

\bibitem{Cardoso:2003qd}
V.~Cardoso, J.~P.~S. Lemos and S.~Yoshida, \emph{{Scalar gravitational perturbations and quasinormal modes in the five-dimensional Schwarzschild black hole}}, \href{https://doi.org/10.1088/1126-6708/2003/12/041}{\emph{JHEP} {\bfseries 12} (2003) 041} [\href{https://arxiv.org/abs/hep-th/0311260}{{\ttfamily hep-th/0311260}}].

\bibitem{Zhidenko:2006rs}
A.~Zhidenko, \emph{{Massive scalar field quasi-normal modes of higher dimensional black holes}}, \href{https://doi.org/10.1103/PhysRevD.74.064017}{\emph{Phys. Rev. D} {\bfseries 74} (2006) 064017} [\href{https://arxiv.org/abs/gr-qc/0607133}{{\ttfamily gr-qc/0607133}}].

\bibitem{Konoplya:2011qq}
R.~A. Konoplya and A.~Zhidenko, \emph{{Quasinormal modes of black holes: From astrophysics to string theory}}, \href{https://doi.org/10.1103/RevModPhys.83.793}{\emph{Rev. Mod. Phys.} {\bfseries 83} (2011) 793} [\href{https://arxiv.org/abs/1102.4014}{{\ttfamily 1102.4014}}].

\bibitem{Calza:2025whq}
M.~Calz{\'a}, \emph{{GrayHawk: A public code for calculating the Gray Body Factors of massless fields around spherically symmetric Black Holes}}, \href{https://doi.org/10.1016/j.dark.2025.101900}{\emph{Phys. Dark Univ.} {\bfseries 48} (2025) 101900} [\href{https://arxiv.org/abs/2502.04041}{{\ttfamily 2502.04041}}].

\end{thebibliography}\endgroup
\end{document}